\documentclass[12pt]{article}
\usepackage{graphics}
\usepackage{amssymb}
\usepackage{amsmath}
\usepackage{upgreek}

\begin{document}

{\noindent \bf Little Rip and Pseudo Rip phenomena from coupled dark energy}

\bigskip

\begin{center}

I. Brevik\footnote{iver.h.brevik@ntnu.no}

\bigskip
Department of Energy and Process Engineering, Norwegian University of Science and Technology, N-7491 Trondheim, Norway

\bigskip
A. V. Timoshkin and Y. Rabochaya\footnote{Also at Department of Physics, University of Trento, Trento, Italy.}

\bigskip

Tomsk State Pedagogical University, 634061 Tomsk, Russia

\bigskip

\bigskip

%Revised version,
%\today

\end{center}

\bigskip
\begin{abstract}

We consider  Little Rip (LR) and  Pseudo Rip (PR) cosmological models with two interacting ideal fluids, corresponding to dark energy and dark matter. The interaction between the dark energy and the dark matter fluid components is described in terms of the parameters in the equations of state  for the LR and PR universes. In contrast to a model containing only a pure dark energy, the presence of the interaction term between the fluid components in the gravitational equations leads to a modification of the equation of state  parameters. The properties of the early universe in this formalism are pointed out.

\end{abstract}

\section{ Introduction}

	In the present paper we  study a cosmological system containing two coupled fluids: a dark energy component with a linear inhomogeneous equation of state (EoS) and a dark matter component with a linear homogeneous EoS. As is known, a dark energy universe shows a peculiar behavior with respect to singularities in the far future, even in cases where it is induced by modified gravity  \cite{nojiri05} (for reviews, see \cite{nojiri07,capozziello09,capozziello11,nojiri11,nojiri13}). Recently, a general review of dark energy was given in Ref.~\cite{bamba12}.
It is of interest to investigate the influence of a coupling between dark energy and dark matter responsible for the accelerated expansion of the universe. Among various different possible models investigated in the literature , we shall focus attention on those that treat dark energy and dark matter as ideal (nonviscous) fluids with an unusual EoS. Very general dark fluid models with an inhomogeneous EoS were introduced in \cite{nojiri05a,nojiri06,brevik04,capozziello06}.
There exist various cosmological scenarios for the evolution of the universe including the Big Rip \cite{caldwell03,nojiri04,nojiri03}, the Little Rip \cite{frampton11,brevik11,frampton12,astashenok12,astashenokarxiv,astashenok12a,nojiriarxiv,makarenkoarxiv},  the Pseudo Rip \cite{frampton12a}  and the Quasi Rip \cite{wei12}.

In the present work we investigate the influence from the interaction between  dark energy and  dark matter via the time-dependent thermodynamic parameter $w$  and the cosmological constant  $\Lambda$.  That means, the EoS will play an important role, for the LR as well as for the PR  phenomena. For the two fluid components we obtain expressions for  the dark energy density, the dark matter density, and the cosmological 'constant', called $\Lambda(t)$.  In particular, we derive the parameters for the early stage of the history of the universe.

\section{Little Rip  models with interaction between dark energy and dark matter}

We consider a universe filled with two interacting ideal fluids components: a dark energy, and a dark matter, in a spatially flat Friedman-Robertson-Walker metric with scale factor  $a$. The background equations are given by \cite{nojiri05}
\begin{align}
& \dot{\rho}+3H(\rho+p)=-Q, \notag \\
& \dot{\rho}_m+3H(\rho_m+p_m)=Q, \notag \\
& \dot{H}=-\frac{1}{2}k^2(\rho+p+\rho_m+p_m), \label{1}
\end{align}
where  $H=\dot{a}/a$ is the Hubble rate and  $k^2=8\pi G$  with  $G$ being Newton's gravitational constant. Moreover $\rho,p$ and $\rho_m, p_m$      are the energy density and the pressure  of a dark energy and a dark matter, respectively, and $Q$    is the interaction term between  dark energy and  dark matter. A dot denotes the derivative with respect to cosmic time $t$.

Friedman's equation for the Hubble rate is
\begin{equation}
H^2=\frac{1}{3}k^2(\rho+\rho_m). \label{2}
\end{equation}
We will now investigate  various cosmological models.

\subsection{Little Rip model: first variant}

	Let us consider a LR model with the following form for Hubble parameter \cite{frampton12}:
\begin{equation}
H=H_0 \exp(\lambda t), \quad H_0>0, \lambda >0. \label{3}
\end{equation}
We will follow the development of the universe beginning at some time $t=0$ which will at first be left unspecified, except that it refers to an initial instant in the very early universe.  Its precise meaning will be dependent on which model we consider.  The symbol  $H_0$ means the Hubble parameter at this particular instant.

Now assume that the dark energy obeys an inhomogeneous EoS \cite{nojiri05a}:
\begin{equation}
p= w(t)\rho +\Lambda(t), \label{4}
\end{equation}
and that the dark matter obeys another inhomogeneous EoS \cite{nojiri05}:
\begin{equation}
p_m=\tilde{w}(t)\rho_m. \label{5}
\end{equation}
The gravitational equation of motion for dark matter can then be written as
\begin{equation}
\dot{\rho}_m+3H(1+\tilde{w})\rho_m=Q. \label{6}
\end{equation}
We will assume that the thermodynamic parameter $\tilde{w}(t)$ for dark matter has the following form:
\begin{equation}
\tilde{w}(t)=-1+\exp{(-\lambda t)}. \label{7}
\end{equation}
It means that $\tilde{w}$ is  equal to zero at the initial instant $t=0$, and is thereafter decreasing asymptotically to $-1$ in the far future.

 As for the interaction term between dark energy, we will take it to depend linearly on time:
 \begin{equation}
 Q=Q_1t+Q_2, \label{8}
 \end{equation}
 where $Q_1$ and $Q_2$ are constants. A motivation for this choice can be  found in modified gravity theory \cite{nojiri07,nojiri11}.

 The solution of equation (\ref{6}) can be written as
 \begin{equation}
 \rho_m(t)=\frac{1}{3H_0}\left[\tilde{Q}_2(1-\exp(-3H_0t))+Q_1t \right], \label{9}
 \end{equation}
 where $\tilde{Q}_2 \equiv Q_2-Q_1/(3H_0)$. If $t\rightarrow 0$, then $\rho_m \rightarrow 0$. We thus see what is
 the physical meaning of the instant $t=0$ in this model: it is the time at which  dark matter starts to appear in the universe. When $t\rightarrow \infty, \rho_m \rightarrow (Q_1/3H_0)t$.

 Taking into account equations (\ref{2})-(\ref{4}) we obtain the gravitational equation of motion for the dark energy,
 \begin{equation}
 \frac{6\lambda H^2}{k^2}-\dot{\rho}_m+3H\left[ (1+w)\left( \frac{3}{k^2}H^2-\rho_m\right)+\Lambda \right]=-Q. \label{10}
 \end{equation}
Let us choose the parameter  $w(t)$ for dark energy in the form
\begin{equation}
w(t)=-1-\frac{1}{3k^2H^2}. \label{11}
\end{equation}
It means that the deviation from the phantom divide $w=-1$ of the dark energy fluid is at maximum at the initial instant $t=0$, and approaches the limit $-1$ asymptotically in the far future. From equation (\ref{10}) we obtain
\begin{equation}
\Lambda(t)=\frac{1}{k^2}\left( \frac{1}{k^2}-2\lambda H\right) -\frac{H_0}{H }\left( 1+\frac{1}{3k^2H H_0}\right) \rho_m. \label{12}
\end{equation}
In particular, at $t=0$ where $w(0) \equiv w_{\rm in}$ and $\Lambda(0)\equiv \Lambda_{\rm in}$ we have
\begin{equation}
w_{\rm in}=-1-\frac{1}{3k^2H_0^2}, \quad \Lambda_{\rm in}=\frac{1}{k^2}\left( \frac{1}{k^2}-2\lambda H_0\right). \label{13}
\end{equation}
When $t\rightarrow \infty$, one sees that $\Lambda(t) \rightarrow -\infty$.

\subsection{Little Rip model: second variant}

Next we will investigate another example of a LR model in which the parameter $H$  equals \cite{frampton12}
\begin{equation}
H(t)=H_0\exp(C\exp(\lambda t)), \label{14}
\end{equation}
where $H_0, C$ and $\lambda$ are positive constants. We start from  $t=0$, as above. The increase of $H(t)$ with time assumed in equation (\ref{14}) is seen to be much stronger than it was in equation (\ref{3}).

For the thermodynamic dark matter parameter $\tilde{w}(t)$ in the EoS (\ref{5}) we assume now the form
\begin{equation}
\tilde{w}(t)=-1+\exp(\lambda t). \label{15}
\end{equation}
Note that $\tilde{w}(t)$ is always zero or positive: it is zero at the initial instant $t=0$; thereafter it increases  exponentially with increasing $t$.

As for the interaction term $Q$ we take it to be proportional to the Hubble parameter $H$ times an exponential of $\lambda t$:
\begin{equation}
Q=\frac{Q_0H}{H_0}\exp(\lambda t), \label{16}
\end{equation}
$Q_0$ being an unspecified constant.

Solving the gravitational equation of motion (\ref{6}) for dark matter, we find
\begin{equation}
\rho_m(t)=\frac{Q_0}{3H_0}\left\{ 1-\exp \left(\frac{3H_0}{C\lambda}\exp \left( C\left[ 1-\exp\left( C[\exp(\lambda t)-1]\right) \right]\right)\right) \right\}. \label{17}
\end{equation}
We take the parameter $w(t)$ for dark energy in equation (\ref{4}) to have the same form (\ref{11}) as before.
Taking into account equations (\ref{11}), (\ref{16}) and (\ref{17}), we then obtain from equation (\ref{10})
\begin{equation}
\Lambda(t)=\frac{1}{k^2}\left[ \frac{1}{k^2}-2C\lambda H\exp(\lambda t)\right]-\left[\exp(\lambda t)+\frac{1}{3k^2H^2}\right] \rho_m. \label{18}
\end{equation}
In the early universe at $t=0$ we get
\begin{equation}
 w_{\rm in}=-1-\frac{1}{3k^2H_0^2 \exp(2C)}, \quad
\Lambda_{\rm in}=\frac{1}{k^2}\left[\frac{1}{k^2}-2C\lambda H_0 \exp(C)\right]. \label{19}
\end{equation}
These expressions are to be compared with expressions (\ref{13}) above.

In contrast to the previous model, there is no special property possessed by $\rho_m$ at the instant $t=0$ in this model. When $t=0$ equation (\ref{17}) simply yields
\begin{equation}
\rho_m(0)=\frac{Q_0}{3H_0}\left\{ 1-\exp \left(\frac{3H_0}{C\lambda}\right) \right\}. \label{20}
\end{equation}
In the far future, $t\rightarrow \infty$, we see moreover that the dark matter gradually fades away, $\rho_m(\infty)=0$. This can be looked upon as a characteristic property analogous to the property $\rho_m(0)=0$ in our previous model. It should be noted that for both models, $p_m(0)=0$.

 When $t\rightarrow \infty$, we again find that $\Lambda(t)\rightarrow -\infty$.

So far we have thus constructed two examples of LR cosmology, in terms of time-dependent parameters in the EoS, taking into account an interaction term $Q$ between dark energy and dark matter.

\section{Pseudo Rip model}

We will now investigate examples in which the Hubble parameter approaches a constant in the far future. That means, the universe approaches asymptotically a de Sitter space. We will make this analysis in analogy with the LR model above.

{\bf First example.} Assume that the Hubble parameter is given as \cite{frampton12}
\begin{equation}
H=H_0-H_1\exp(-\lambda t), \label{21}
\end{equation}
where $H_0, H_1$ and $\lambda$ are positive constants, $H_0>H_1$, and $t>0$.

Let us suppose that the parameter  $\tilde{w}(t)$ for dark matter in (\ref{5}) has the form (\ref{15}). Choosing the  interaction between dark energy and dark matter in the form
\begin{equation}
Q=\frac{3Q_0}{H_0}H\exp(\lambda t), \label{22}
\end{equation}
we find that the solution of the gravitational equation (\ref{6}) is given by
\begin{equation}
\rho_m(t)=\frac{Q_0}{H_0}\left\{1-\exp\left[ -3\left( \frac{H_0}{\lambda}(\exp(\lambda t)-1)\right)-H_1t\right] \right\}. \label{23}
\end{equation}
According to this expression the dark matter energy density is thus $\rho_m=0$ at $t=0$ and  $\rho_m=1$ at $t=\infty$.

If the thermodynamic parameter $w(t)$ for dark energy has the form (\ref{11}), we obtain from (\ref{10}) the following expression for the cosmological 'constant':
\begin{equation}
\Lambda(t)=\frac{1}{k^2}\left[ \frac{1}{k^2}-2\lambda H_1\exp(-\lambda t)\right]+(w-\tilde{w})\rho_m. \label{24}
\end{equation}
In the early universe $t=0$ we have
\begin{equation}
w_{\rm in}=-1-\frac{1}{3k^2(H_0-H_1)^2}, \quad \Lambda_{\rm in}=\frac{1}{k^2}\left(\frac{1}{k^2}-2\lambda H_1\right). \label{25}
\end{equation}

{\bf Second example.} This example considers a cosmological model with asymptotically de Sitter evolution \cite{astashenok12a}. The Hubble parameter is taken equal to \cite{brevik13}
\begin{equation}
H=\frac{x_f}{\sqrt 3}\left[1-\left(1-\frac{x_0}{x_f}\right)\exp\left(-\frac{\sqrt{3}At}{2x_f}\right) \right], \label{26}
\end{equation}
where $x_0=\sqrt{\rho_0}$ is the energy density at present, $x_f$ is a finite value related to divergence of the  cosmological time \cite{brevik13}, and $A$  is a positive constant. When $t\rightarrow \infty$ the Hubble parameter $H\rightarrow x_f/\sqrt 3$ and the expression (\ref{26}) asymptotically tends to the de Sitter solution.

Let us take the parameter $\tilde{w}(t)$ in the form
\begin{equation}
\tilde{w}(t)=-1+\exp\left(\frac{\sqrt{3}At}{2x_f}\right), \label{27}
\end{equation}
and take the interaction term in the form
\begin{equation}
Q=Q_0\exp \left(-\frac{2x_f^2}{A}\exp \left(\frac{\sqrt{3}At}{2x_f}\right)\right), \label{28}
\end{equation}
where $Q_0$ is a constant.

In this case the solution of the gravitational equation (\ref{6}) for dark matter becomes
\begin{equation}
\rho_m(t)=\frac{Q}{\sqrt{3}(x_f-x_0)}\left[\exp(\sqrt{3}(x_f-x_0)t)-1\right]. \label{29}
\end{equation}
using (\ref{6}), (\ref{11}) and (\ref{29}) to solve (\ref{10}) with respect to $\Lambda(t)$, we obtain
\begin{equation}
\Lambda(t)=\frac{1}{k^2}\left[ \frac{1}{k^2}-A\left(1-\frac{\sqrt 3}{x_f}H\right)\right]-\left[ \frac{1}{3k^2H^2}+\frac{x_f-x_0}{x_f-\sqrt{3}H}\right] \rho_m. \label{30}
\end{equation}
In the early universe one gets
\begin{equation}
w_{\rm in}=-1-\frac{1}{3k^2 x_0^2}, \quad \Lambda_{\rm in}=\frac{1}{k^2}\left[ \frac{1}{k^2}-A\left(1-\frac{x_0}{x_f}\right)\right]. \label{31}
\end{equation}
Thus we have investigated, via the parameters of the EoS, the influence of the dark energy/dark matter coupling on the evolution of the Pseudo Rip model.

\section{Conclusion}

In the present work we studied  Little Rip and Pseudo Rip cosmological models taking into account the coupling between dark energy and dark matter. The gravitational equations of motion for dark matter were solved, and the descriptions of the LR and PR universes in terms of the EoS parameters were given. Unlike a model containing only  a pure dark energy, the presence of an interaction term between dark energy and dark matter in the gravitational equations leads to corrections in the EoS parameters; see Refs.~\cite{brevik13,brevik12,brevik13a,brevik13b}. The case of the very early universe was emphasized.

We mention finally that there exist also other investigations dealing with the coupling between the dark energy and dark matter components,  from various perspectives; cf., for instance,  the Archimedean-type approach followed by Balakin and Bochkarev \cite{balakin11,balakin11a}.

\bigskip

{\bf Acknowledgement}

\bigskip

This work has been supported by project 2.1839.2011 of Ministry of Education and Science (Russia) and LRSS project 224.2012.2 (Russia). We thank Professor Sergei Odintsov for very useful discussions and valuable remarks.

%\newpage

\end{document}